\documentclass[conference,a4paper]{IEEEtran}

\usepackage{amsmath}%
\usepackage{amsfonts}%
\usepackage{amssymb}%
\usepackage{graphicx}

\usepackage{pdflscape}
\usepackage{multirow}
\newcommand{\Ts}{T_{\text{symbol}}} 

\newcommand{\blk}{n}		

\newcommand{\Y}{\textbf{Y}}

\newcommand{\SNR}{\textsf{SNR}}

\usepackage{pgf}
\usepackage{tikz}
\usetikzlibrary{arrows,automata}
\usepackage[latin1]{inputenc}
\usepackage{verbatim}

\begin{document}

\sloppy

\title{On Wiener Phase Noise Channels\\ at High Signal-to-Noise Ratio}

\author{
\IEEEauthorblockN{Hassan Ghozlan}
\IEEEauthorblockA{Department of Electrical Engineering\\
University of Southern California\\
Los Angeles, CA 90089 USA\\
ghozlan@usc.edu}
\and
\IEEEauthorblockN{Gerhard Kramer}
\IEEEauthorblockA{
Institute for Communications Engineering \\
Technische Universit\"{a}t M\"{u}nchen \\
80333 Munich, Germany \\
gerhard.kramer@tum.de}
}



\maketitle

\begin{abstract}
Consider a waveform channel where the transmitted signal is corrupted by Wiener phase noise
and additive white Gaussian noise (AWGN).
A discrete-time channel model that takes into account the effect of filtering on the phase noise is developed.
The model is based on a multi-sample receiver
which, at high Signal-to-Noise Ratio (SNR), achieves a rate that grows logarithmically with the SNR
if the number of samples per symbol grows with the square-root of the SNR.
Moreover, the pre-log factor is at least 1/2 in this case.
\end{abstract}

\section{Introduction}
Phase noise is an impairment that often arises in coherent communication systems.
Different models are adopted for the phase noise process depending on the application.
In \cite{Katz2004}, Katz and Shamai studied a discrete-time model of a phase noise channel (partially coherent channel)
in which the phase noise is independent and identically distributed (i.i.d.) with a Tikhonov distribution.
This model is reasonable for the residual phase error of a phase-tracking scheme, such as a Phase-Locked Loop (PLL).
In \cite{Goebel2011}, the authors investigate white (Gaussian) phase noise for which they observed a ``spectral loss'' phenomenon.
The white phase noise approximates the nonlinear effect of cross-phase modulation (XPM) in a Wavelength-Division Multiplexing (WDM) optical communication system.
Lapidoth studied in \cite{LapidothPhaseNoise2002} a \emph{discrete-time} phase noise channel 
\begin{align}
 Y_k = X_k e^{j \Theta_k} + N_k
\label{eq:dt-pnch-lapidoth}
\end{align}
at high SNR, 
where $\{Y_k\}$ is the output, $\{X_k\}$ is the input, $\{\Theta_k\}$ is the phase noise process and $\{N_k\}$ is the additive noise.
He considered both memoryless phase noise and phase noise with memory. 
He showed that the capacity grows \emph{logarithmically} with the SNR with a pre-log factor 1/2,
where the pre-log is due to amplitude modulation only.
The phase modulation contributes a bounded number of bits only.

In this paper, we study a communication system in which the transmitted \emph{waveform} is corrupted 
by Wiener phase noise and AWGN. The model is  
\begin{align}
  r(t) = x(t) \ \exp(j \theta(t)) + n(t),
  \text{ for } t \in \mathbb{R}
  \label{eq:waveform_ch}
\end{align}
where $x(t)$ and $r(t)$ are the transmitted and received signals, respectively,
while $n(t)$ and $\theta(t)$ are the additive and phase noise, respectively.
A detailed description of the model is given in Sec. \ref{sec:ct-model}.
One application for such a channel model is optical communication under linear propagation,
in which the laser phase noise is a continuous-time Wiener process
(see \cite{Foschini1988IT} and references therein).
Since the sampling of a continuous-time Wiener process yields a discrete-time Wiener process (Gaussian random walk),
it is tempting to use the model (\ref{eq:dt-pnch-lapidoth})
with $\{\Theta\}$ as a discrete-time Wiener process,
but this ignores the effect of \emph{filtering} prior to sampling.
It was pointed out in \cite{Foschini1988IT} that 
``even coherent systems relying on amplitude modulation 
(phase noise is obviously a problem in systems employing phase modulation) 
will suffer some degradation due to the presence of phase noise''.
This is because the filtering converts phase fluctuations to amplitude variations.
It is worth mentioning that filtering is necessary before sampling to limit the variance of the noise samples.

The model (\ref{eq:dt-pnch-lapidoth}) thus does not fit the channel (\ref{eq:waveform_ch})
and it is not obvious whether a pre-log 1/2 is achievable.
The model that takes the effect of (matched) filtering into account is
\begin{align}
 Y_k = X_k H_k + N_k
 \label{eq:dt-noncoherent-fade}
\end{align}
where $\{H_k\}$ is a fading process.
The model (\ref{eq:dt-noncoherent-fade}) falls in the class of non-coherent fading channels,
i.e., the transmitter and receiver have knowledge of the distribution of the fading process $\{H_k\}$, 
but have no knowledge of its realization.
For such channels, Lapidoth and Moser showed in \cite{Lapidoth2003} that, 
at high SNR, the capacity grows \emph{double-logarithmically} with the SNR, 
when the process $\{H_k\}$ is stationary, ergodic, and \emph{regular}.

Rather than using a matched filter and sampling its output at the symbol rate,
we use a multi-sample receiver, i.e., a filter whose output is sampled many times per symbol.
We show that this receiver achieves a rate that grows \emph{logarithmically} with the SNR
if the number of samples per symbol grows with the square-root of the SNR.
Furthermore, we show that a pre-log of 1/2 is achievable through amplitude modulation.
In this paper, we study only rectangular pulses
but we believe that the results hold qualitatively for other pulses.

The paper is organized as follows.
The continuous-time model is described in Sec. \ref{sec:ct-model}
and the discretization is described in Sec. \ref{sec:dt-model}.
We derive a lower bound on the capacity in Sec. \ref{sec:amplitude-lower-bound}
and discuss our result in Sec. \ref{sec:discuss}.
Finally, we conclude the paper with Sec. \ref{sec:conc}.

\section{Continuous-Time Model}
\label{sec:ct-model}
We use the following notation: 
$j=\sqrt{-1}$ , 
$^*$ denotes the complex conjugate, 
$\delta_D$ is the Dirac delta function, 
$\lceil \cdot \rceil$ is the ceiling operator,
$\Re[\cdot]$ is the real part of a complex number,
$\log(\cdot)$ is the natural logarithm and
we use $X_1^k$ to denote the $k$-tuple $(X_1,X_2,\ldots,X_k)$.
Suppose the transmit-waveform is $x(t)$ and 
the receiver observes
\begin{align}
  r(t) = x(t) \ \exp(j \theta(t)) + n(t)
\end{align}
where $n(t)$ is a realization of a white circularly-symmetric complex Gaussian process $N(t)$ with
\begin{align}
&\mathbb{E}\left[ N(t) \right] = 0 \nonumber \\
&\mathbb{E}\left[ N(t_1) N^*(t_2) \right] = \sigma^2_N ~ \delta_D(t_2-t_1).
\label{eq:Zt_statistics}
\end{align}
The phase $\theta(t)$ is a realization of a Wiener process $\Theta(t)$:
\begin{align}
  \Theta(t) = \Theta(0) + \int_0^t W(\tau) d\tau
\label{eq:Thetat}
\end{align}
where $\Theta(0)$ is uniform  on $[-\pi,\pi)$ and 
$W(t)$ is a real Gaussian process with
\begin{align}
&\mathbb{E}\left[ W(t) \right] = 0 \\
&\mathbb{E}\left[ W(t_1) W(t_2)\right] = 2\pi \beta ~ \delta_D(t_2-t_1) .
\label{eq:Wt_statistics}
\end{align}
The processes $N(t)$ and $\Theta(t)$ are independent of each other and independent of the input as well.
$N_0 = 2 \sigma^2_N$ is the single-sided power spectral density of the additive noise.
The parameter $\beta$ is called the full-width at half-maximum (FWHM),
because the power spectral density of $e^{j\Theta(t)}$ has a Lorentzian shape,
for which $\beta$ is the full-width at half the maximum.
The transmitted waveforms must satisfy the power constraint
\begin{align}
	\mathbb{E}\left[\frac{1}{T} \int_0^{T} |X(t)|^2 dt \right] \leq \mathcal{P}
	\label{eq:finitesupport_waveform_power_constraint}
\end{align}
where $T$ is the transmission interval.

\section{Discrete-Time Model}
\label{sec:dt-model}
Let $(x_1,x_2,\ldots,x_n)$
be the codeword sent by the transmitter.
Suppose the transmitter uses a unit-energy rectangular pulse,
i.e., the waveform sent by the transmitter is
\begin{align}
 x(t) = \sum_{m=1}^{\blk}  x_m \ g(t-(m-1) \Ts)
\label{eq:xt_modulated_rect}
\end{align}
where $\Ts$ is the symbol interval and
\begin{align}
 g(t) \equiv \left\{ 
  \begin{array}{ll}
  \sqrt{1/\Ts},	& 0 \leq t <\Ts, \\
  0,			& \text{otherwise}.
  \end{array}
 \right.
\label{eq:rect_pulse_def}
\end{align}

Let $L$ be the number of samples per symbol ($L \geq 1$)
and define the sample interval $\Delta$ as
\begin{align}
\Delta = \frac{\Ts}{L}.
\end{align}

The received waveform $r(t)$ is filtered using an integrator over a sample interval to give the output signal
\begin{align}
y(t) 
&= \int_{t - \Delta}^{t} r(\tau) \ d\tau.
\end{align}
where $y(t)$ is a realization of $Y(t)$.
The output $Y(t)$ is sampled every $\Delta$ seconds
which yields the discrete-time model:
\begin{align}
Y_k = X_{\lceil k/L \rceil} \Delta \ e^{j \Theta_k} \ F_k + N_k
\label{eq:Yk}
\end{align}
for $k=1,\ldots,n L$,
where
$Y_k \equiv Y(k \Delta)$, 
$\Theta_k \equiv \Theta( (k-1) \Delta )$,
\begin{align}
F_k \equiv \frac{1}{\Delta} \int_{(k-1) \Delta}^{k \Delta} e^{j(\Theta(\tau)-\Theta_k)} \ d\tau
\label{eq:Fk_def}
\end{align}
and
\begin{align}
N_k 
&\equiv \int_{(k-1) \Delta}^{k \Delta} N(\tau) \ d\tau.
\label{eq:Nk_def}
\end{align}
The process $\{N_k\}$ is an i.i.d. circularly-symmetric complex Gaussian process with mean $0$ and 
$\mathbb{E}[ |N_k|^2 ] = \sigma^2_N \Delta$
while the process $\{\Theta_k\}$ is the discrete-time Wiener process:
\begin{align}
  \Theta_k = \Theta_{k-1} + W_k 
\end{align}
where
$\Theta_1$ is uniform on $[-\pi,\pi)$ and
$\{W_k\}$ is an i.i.d. real Gaussian process with mean $0$ and $\mathbb{E}[ |W_k|^2 ] = 2\pi \beta \Delta$.
The process $\{F_k\}$ is an i.i.d. process.
Moreover, $\{F_k\}$ and $\{W_k\}$ are independent of $\{N_k\}$ but not independent of each other.

Equations (\ref{eq:finitesupport_waveform_power_constraint}) -- (\ref{eq:rect_pulse_def})
imply the power constraint 
\begin{align}
	\frac{1}{\blk} \sum_{m=1}^{\blk} \mathbb{E}[|X_m|^2] \leq P = \mathcal{P} \Ts.
	\label{eq:dt_power_constraint}
\end{align}

\section{Lower Bound}
\label{sec:amplitude-lower-bound}
For the $k$th input symbol $X_k$ we have $L$ outputs,
so it is convenient to group the $L$ samples per symbol in one vector and define
$\Y_k \equiv (Y_{(k-1) L + 1},Y_{(k-1) L + 2},\ldots,Y_{(k-1) L + L})$.
We further define $X_A \equiv |X|$ and $X_{\Phi} \equiv \angle X$.
We decompose the mutual information using the chain rule into two parts:
\begin{align}
I(X_1^n;\Y_1^n) 
&= I(X_{A,1}^n;\Y_1^n) + I(X_{\Phi,1}^n;\Y_1^n|X_{A,1}^n).
\label{eq:I_X1n_Y1n}
\end{align}
The first term represents the contribution of the amplitude modulation while 
the second term represents the contribution of the phase modulation.
We focus on the amplitude contribution 
and use $I(X_{\Phi,1}^n;\Y_1^n|X_{A,1}^n) \geq 0$ to obtain the lower bound
\begin{align}
I(X_1^n;\Y_1^n) 
&\geq I(X_{A,1}^n;\Y_1^n).
\label{eq:I_X1n_Y1n_LB}
\end{align}

Suppose that $X_{A,1}^n$ is i.i.d.
Hence, we have
\begin{align}
I(X_{A,1}^n;\Y_1^n)
&\stackrel{(a)}{=} \sum_{k=1}^n I(X_{A,k};\Y_1^n|X_{A,1}^{k-1}) \nonumber \\
&\stackrel{(b)}{=} \sum_{k=1}^n H(X_{A,k}) - H(X_{A,k}|\Y_1^n ~ X_{A,1}^{k-1}) \nonumber \\
&\stackrel{(c)}{\geq} \sum_{k=1}^n I(X_{A,k};\Y_k) \nonumber \\
&\stackrel{(d)}{\geq} \sum_{k=1}^n I(X_{A,k};V_k)
\label{eq:I_XA1n_Y1n_LB}
\end{align}
where
\begin{align}
V_k = \sum_{\ell=1}^L |Y_{(k-1)L+\ell}|^2.
\label{eq:V_def}
\end{align}
Step
$(a)$ follows from the chain rule of mutual information, 
$(b)$ follows from the independence of $X_{A,1},X_{A,2},\ldots,X_{A,n}$, 
$(c)$ holds because conditioning does not increase entropy, and
$(d)$ follows from the data processing inequality.
Since $X_{A,1}^n$ is identically distributed, then $V_1^n$ is also identically distributed and 
we have, for $k \geq 2$,
\begin{align}
I(X_{A,k};V_k) = I(X_{A,1};V_1).
\label{eq:I_XAk_Vk}
\end{align}

In the rest of this section, we consider only one symbol ($k=1$) and drop the time index.
Moreover, we assume that $\Ts=1$ for simplicity.
By combining (\ref{eq:V_def}) and (\ref{eq:Yk}), we have
\begin{align}
V 
&= \sum_{\ell=1}^L \left( X_A^2 \Delta^2 |F_\ell|^2 + 2 X_A \Delta \Re[ e^{j \Phi_X} e^{j \Theta_\ell} F_\ell N_\ell^* ] + |N_\ell|^2 \right) \nonumber \\
&= X_A^2 \Delta G+ 2 X_A \Delta Z_1 + Z_0
\label{eq:V_direct}
\end{align}
where $G$, $Z_1$ and $Z_0$ are defined as
\begin{align}
G &\equiv \frac{1}{L} \sum_{\ell=1}^L |F_\ell|^2
\label{eq:G_def} \\
Z_1 &\equiv \sum_{\ell=1}^L \Re[ e^{j \Phi_X} e^{j \Theta_\ell} F_\ell N_\ell^* ]
\label{eq:Z1_def} \\
Z_0 &\equiv \sum_{\ell=1}^L |N_\ell|^2.
\label{eq:Z0_def} 
\end{align}
The second-order statistics of $Z_1$ and $Z_0$ are
\begin{align}
\begin{array}{ll}
\mathbb{E}[ Z_1 ] = 0 		& \textsf{Var}[ Z_1 ] = \mathbb{E}[G] {\sigma^2_N}/{2} \\
\mathbb{E}[ Z_0 ] = \sigma^2_N 	& \textsf{Var}\left[ Z_0 \right] = \sigma^4_N \Delta \\
\mathbb{E}\left[ Z_1 (Z_0-\mathbb{E}[Z_0]) \right] = 0.
\end{array}
\label{eq:second-order-stats}
\end{align}

By using the Auxiliary-Channel Lower Bound Theorem in \cite[Sec. VI]{Arnold2006}, we have
\begin{align}
I(X_A;V)
&\geq \mathbb{E}[-\log{Q_V(V)}] + \mathbb{E}[\log{Q_{V|X_A}(V|X_A)}]
\label{eq:I_XA_V_LB_AuxCh}
\end{align}
where $Q_{V|X_A}(v|x_A)$ is an arbitrary auxiliary channel and
\begin{align}
Q_V(v) = \int P_{X_A}(x_A) Q_{V|X_A}(v|x_A) dx_A
\label{eq:QV_def}
\end{align}
where $P_{X_A}(\cdot)$ is the \emph{true} input distribution,
i.e.,
$Q_V(\cdot)$ is the output distribution obtained by connecting the true input source to the auxiliary channel.
$\mathbb{E}[\cdot]$ is the expectation according to the \emph{true} distribution.
We choose the auxiliary channel
\begin{align}
Q_{V|X_A}(v|x_A) 
= \frac{1}{\sqrt{4 \pi x_A^2 \Delta^2 \sigma^2_N}}
\exp\left(- \frac{(v-x_A^2 \Delta - \sigma^2_N)^2}{4 x_A^2 \Delta^2 \sigma^2_N} \right).
\label{eq:Q_V|XA}
\end{align}
It follows that
\begin{align}
&\mathbb{E}[ -\log( Q_{V|X_A}(V|X_A) ) ]
= \mathbb{E}\left[ \frac{(V-X_A^2 \Delta - \sigma^2_N)^2}{4 X_A^2 \Delta^2 \sigma^2_N} \right] \nonumber \\& \qquad \qquad
+ \log{\Delta} 
+ \frac{1}{2} \log(4 \pi\sigma^2_N)
+ \frac{1}{2} \mathbb{E}[ \log(X_A^2) ].
\label{eq:log_Q_V|XA_original}
\end{align}
By using (\ref{eq:V_direct}), we have
\begin{align}
&\left( V-X_A^2 \Delta  - \sigma^2_N \right)^2 \nonumber\\
&= \left( X_A^2 \Delta (G-1) + 2 X_A \Delta Z_1 + (Z_0-\sigma^2_N) \right)^2 \nonumber\\
&=X_A^4 \Delta^2 (G-1)^2
+ 4 X_A^2 \Delta^2 Z_1^2 
+ (Z_0-\sigma^2_N)^2 \nonumber\\&
+ 4 X_A^3 \Delta^2 (G-1) Z_1
+ 2 X_A^2 \Delta (G-1) (Z_0-\sigma^2_N) \nonumber \\&
+ 4 X_A \Delta Z_1 (Z_0-\sigma^2_N)
\end{align}
and hence, using the second-order statistics (\ref{eq:second-order-stats}), we have
\begin{align}
&\mathbb{E}\left[ \frac{(V-X_A^2 \Delta - \sigma^2_N)^2}{4 X_A^2 \Delta^2 \sigma^2_N} \right] \nonumber\\
&=\frac{1}{4 \sigma^2_N} P \ \mathbb{E}\left[ (G-1)^2 \right] 
+ \frac{1}{2} \mathbb{E}[G]
+ \frac{\sigma^2_N}{4 \Delta} \mathbb{E}\left[ \frac{1}{X_A^2} \right]
\label{eq:big_expectation}
\end{align}
where we also used
\begin{align}
 \mathbb{E}\left[ (G - 1) Z_1 \right] = 0.
\end{align}
Substituting (\ref{eq:big_expectation}) into (\ref{eq:log_Q_V|XA_original}) and using $\mathbb{E}[G] \leq 1$ yield
\begin{align}
&\mathbb{E}[ -\log( Q_{V|X_A}(V|X_A) ) ] \nonumber \\
&\leq \log{\Delta}
+ \frac{1}{2} \log(4 \pi\sigma^2_N)
+ \frac{1}{2} \mathbb{E}[ \log(X_A^2) ] \nonumber \\&
+ \frac{P}{4 \sigma^2_N} \mathbb{E}\left[ (G-1)^2 \right]  + \frac{1}{2} + \frac{\sigma^2_N}{4 \Delta} \mathbb{E}\left[ \frac{1}{X_A^2} \right].
\label{eq:log_Q_V|XA}
\end{align}

It is convenient to define $X_P \equiv X_A^2$.
We choose the input distribution
\begin{align}
P_{X_P}(x_P) = \left\{ 
  \begin{array}{ll}
  \frac{1}{\lambda} \exp\left(-\frac{x_P-P_{\min}}{\lambda}\right),	& x_P \geq P_{\min} \\
  0,			& \text{otherwise}
  \end{array}
 \right.
\label{eq:P_XP}
\end{align}
where $0 < P_{\min} < P$ and $\lambda = P - P_{\min}$,
so that
\begin{align}
\mathbb{E}[X_P] = \mathbb{E}[X_A^2] = P.
\label{eq:E_XP}
\end{align}
It follows from  (\ref{eq:QV_def}) and (\ref{eq:P_XP}) that
\begin{align}
Q_V(v) 
&= \int_{P_{\min}}^\infty \frac{1}{\lambda} \exp\left(-\frac{x_P-P_{\min}}{\lambda}\right) ~ Q_{V|X_P}(v|x_P) ~ dx_P \nonumber \\
&\leq \exp( P_{\min}/\lambda ) ~ F_{V}(v)
\label{eq:QV_FV_ineq}
\end{align}
where
\begin{align}
Q_{V|X_P}(v|x_P) = Q_{V|X_A}(v|\sqrt{x_P})
\label{eq:Q_V|XP}
\end{align}
and
\begin{align}
F_V(v)
&\equiv \int_{0}^\infty \frac{1}{\lambda} \exp\left(-\frac{x_P}{\lambda}\right) \ Q_{V|X_P}(v|x_P) dx_P.
\label{eq:F_V_def}
\end{align}
The inequality $(\ref{eq:QV_FV_ineq})$ follows from the non-negativity of the integrand.
By combining (\ref{eq:Q_V|XA}), (\ref{eq:Q_V|XP}), (\ref{eq:F_V_def}) and making the change of variables $x=x_P \Delta$, we have
\begin{align}
&F_V(v) \nonumber\\
&= \int_0^\infty \frac{e^{-{x}/(\lambda \Delta)}}{\lambda \Delta}
\frac{1}{\sqrt{4 \pi x \Delta \sigma^2_N}} \exp\left(- \frac{(v-x - \sigma^2_N)^2}{4 x \Delta \sigma^2_N} \right) dx \nonumber \\
&= \frac{1}{\sqrt{\lambda \Delta (\lambda \Delta + 4 \Delta \sigma^2_N)}} \times \nonumber\\& \quad
\exp\left( \frac{2}{4 \Delta \sigma^2_N} \left[ v-\sigma^2_N - |v-\sigma^2_N| \sqrt{1+\frac{4 \Delta \sigma^2_N}{\lambda \Delta}} \right] \right)
\label{eq:F_V}
\end{align}
where we used equation (140) in Appendix A of \cite{Moser2012}:
\begin{align}
&\int_0^\infty \frac{1}{a} \exp\left(-\frac{x}{a}\right)
\frac{1}{\sqrt{\pi b x}} \exp\left(- \frac{(u-x)^2}{b x} \right) dx \nonumber\\&
= \frac{1}{\sqrt{a (a + b)}}
\exp\left( \frac{2}{b} \left[ u - |u| \sqrt{1+\frac{b}{a}} \right] \right).
\end{align}
Therefore, we have
\begin{align}
&\mathbb{E}[ -\log( F_{V}(V) ) ] \nonumber \\
&= \frac{1}{2} \log(\Delta^2 (\lambda^2 + 4 \lambda \sigma^2_N)) \nonumber \\&
- \frac{1}{2 \Delta \sigma^2_N} \left[ \mathbb{E}[V-\sigma^2_N] - \mathbb{E}[|V-\sigma^2_N|] \sqrt{1+\frac{4 \sigma^2_N}{\lambda}} \right] \nonumber \\
&\stackrel{(a)}{\geq} \log(\Delta \lambda) + 
\frac{1}{2 \sigma^2_N \Delta} \mathbb{E}[V-\sigma^2_N] \left[ \sqrt{1+\frac{4 \sigma^2_N}{\lambda}} - 1 \right] \nonumber \\
&\stackrel{(b)}{\geq} \log(\Delta \lambda)
\label{eq:log_F_V}
\end{align}
where
$(a)$ holds because the logarithmic function is monotonic and $\mathbb{E}[| \cdot |] \geq \mathbb{E}[ \cdot ]$, and
$(b)$ holds because
\begin{align}
&\mathbb{E}[V-\sigma^2_N] \nonumber \\
&= \mathbb{E}[X_A^2] \Delta \mathbb{E}[G] + 2 \mathbb{E}[X_A] \Delta \mathbb{E}[Z_1] + \mathbb{E}[Z_0] - \sigma^2_N \nonumber \\
&= P \Delta \mathbb{E}[G] \geq 0.
\end{align}
The monotonicity of the logarithmic function and (\ref{eq:QV_FV_ineq}) yield
\begin{align}
\mathbb{E}[ -\log( Q_{V}(V) ) ] 
&\geq \mathbb{E}\left[ -\log\left( e^{P_{\min}/\lambda} F_{V}(V) \right) \right] \nonumber \\
&\geq \log{\Delta} + \log{\lambda}
- \frac{P_{\min}}{\lambda} 
\label{eq:log_Q_V}
\end{align}
where the last inequality follows from (\ref{eq:log_F_V}).
It follows from (\ref{eq:I_XA_V_LB_AuxCh}), (\ref{eq:log_Q_V|XA}) and (\ref{eq:log_Q_V}) that
\begin{align}
I(X_A;V)
&\geq 
\log{\lambda}
- \frac{P_{\min}}{\lambda} 
- \frac{1}{2} \log(4 \pi\sigma^2_N)
- \frac{1}{2} \mathbb{E}[ \log(X_A^2) ] \nonumber \\&
- \frac{P}{4 \sigma^2_N} \mathbb{E}\left[ (G-1)^2 \right] - \frac{1}{2} - \frac{\sigma^2_N}{4 \Delta} \mathbb{E}\left[ \frac{1}{X_A^2} \right].
\label{eq:I_XA_V_LB}
\end{align}

If $P_{\min} = P/2$, then $\lambda = P - P_{\min} = P/2$ and we have 
\begin{align}
\mathbb{E}\left[ \frac{1}{X_P} \right] 
\leq \frac{1}{P_{\min}}
= \frac{2}{P}
\end{align}
and
\begin{align}
\mathbb{E}\left[ \log\left( X_P \right) \right]
&= \int_{\lambda}^\infty \frac{1}{\lambda} e^{-(x-\lambda)/\lambda} \log(x) dx \nonumber \\
&\stackrel{(a)}{=} \log\lambda + \int_{1}^\infty e^{-(u-1)} \log(u) du \nonumber \\
&\stackrel{(b)}{\leq} \log\lambda + 1
\end{align}
where 
$(a)$ follows by the change of variables $u = x/\lambda$, and
$(b)$ holds because $\log(u) \leq u-1$ for all $u > 0$.
Substituting into (\ref{eq:I_XA_V_LB}), we obtain
\begin{align}
I(X_A;V) - \frac{1}{2} \log{\SNR}
&\geq - 2 - \frac{1}{2} \log(8 \pi) 
- \frac{1}{2 \SNR \Delta} \nonumber\\&
- \frac{1}{4} \SNR ~ \mathbb{E}\left[ (G-1)^2 \right]
\end{align}
where $\SNR = {P}/{\sigma^2_N}$.
Suppose $L$
grows with $\SNR$ such that 
\begin{align}
 L = \left\lceil \beta \sqrt{\SNR} \right\rceil.
\end{align}
Since $\Delta = 1/L$, then we have
\begin{align}
\lim_{\SNR \rightarrow \infty} {\SNR \Delta} = \infty
\text{ and }
\lim_{\SNR \rightarrow \infty} \SNR \Delta^2 = \frac{1}{\beta^2}
\end{align}
which implies
\begin{align}
\lim_{\SNR \rightarrow \infty} I(X_A;V) - \frac{1}{2} \log{\SNR}
&\geq - 2  - \frac{1}{2} \log(8 \pi)
- \frac{\pi^2}{36}
\label{eq:I_XA_V_LB_limit}
\end{align}
because (see Appendix)
\begin{align}
\lim_{\Delta \rightarrow 0} \frac{\mathbb{E}[ \left(G - 1 \right)^2 ]}{\Delta^2}
= \frac{(\pi \beta)^2}{9}.
\label{eq:limit_E(G-1)2}
\end{align}

By combining (\ref{eq:I_X1n_Y1n_LB}), (\ref{eq:I_XA1n_Y1n_LB}), (\ref{eq:I_XAk_Vk}) and (\ref{eq:I_XA_V_LB_limit}), we have
\begin{align}
\lim_{\SNR \rightarrow \infty} \frac{1}{n} I(X_1^n;\Y_1^n) - \frac{1}{2} \log{\SNR}
\geq - 2  - \frac{1}{2} \log(8 \pi) - \frac{\pi^2}{36}.
\end{align}
This shows that the information rate grows logarithmically at high SNR with a pre-log factor of 1/2.

\section{Discussion}
\label{sec:discuss}
There is a wide literature on the design of receivers for the channel model 
(\ref{eq:dt-pnch-lapidoth})
with a discrete-time Wiener phase noise,
e.g., see \cite{Barbieri2007}, \cite{Spalvieri2011}, \cite{Barbieri2011} and references therein.
One may want to make use of these designs, which raises the following question:
``when is it justified to approximate the non-coherent fading model (\ref{eq:dt-noncoherent-fade}) with the discrete-time phase noise model (\ref{eq:dt-pnch-lapidoth})?''
Our result suggests that this approximation may be justified 
when the phase variation is small over one symbol interval 
(i.e., when the phase noise linewidth is small compared to the symbol rate) 
\emph{and} also the SNR is low to moderate. 
It must be noted that the SNR at which the high-SNR asymptotics start to manifest themselves depends on the application. 

We remark that
the authors of \cite{Foschini1988Comm} treated on-off keying transmission in the presence of Wiener phase noise
by using a double-filtering receiver, which is composed of
an intermediate frequency (IF) filter, followed by an envelope detector (square-law device) and then a post-detection filter.
They showed that by optimizing the IF receiver bandwidth the double-filtering receiver outperforms the single-filtering (matched filter) receiver.
Furthermore, they showed via computer simulation that the optimum IF bandwidth increases with the SNR.
This is similar to our result in the sense that
we require the number of samples per symbol to increase with the SNR 
in order to achieve a rate that grows logarithmically with the SNR.

Finally, we remark that we have not computed the contribution of phase modulation to the information rate.
We believe that using the multi-sample receiver it is possible to achieve an overall pre-log that is larger than 1/2.
This matter is currently under investigation.

\section{Conclusion}
\label{sec:conc}
We studied a communication system impaired by Wiener phase noise and AWGN.
A discrete-time channel model based on filtering and oversampling is considered.
The model accounts for the filtering effects on the phase noise.
It is shown that at high SNR the multi-sample receiver achieves rates that grow logarithmically 
with at least a 1/2 pre-log factor
if the number of samples per symbol grows with the square-root of the SNR.

\section*{Acknowledgment}
H. Ghozlan was supported by a USC Annenberg Fellowship and NSF Grant CCF-09-05235.
G. Kramer was supported by an Alexander von Humboldt Professorship endowed by
the German Federal Ministry of Education and Research.

\section*{Appendix}
We discuss the limit in (\ref{eq:limit_E(G-1)2}).
We express $\mathbb{E}[ \left(G - 1\right)^2 ]$ as
\begin{align}
\mathbb{E}[ \left(G - 1\right)^2 ]
&= \textsf{Var}(G)
+ \left(\mathbb{E}[G] - 1\right)^2 \nonumber \\
&= \frac{1}{L} \textsf{Var}(|F_1|^2)
+ \left(\mathbb{E}[ |F_1|^2 ] - 1\right)^2
\label{eq:mean_square_G-1}
\end{align}
where the last equality follows from the definition of $G$ in (\ref{eq:G_def}) and because $\{F_k\}$ is i.i.d.

Next, we outline the steps for computing $\mathbb{E}[ |F_1|^4 ]$ and $\mathbb{E}[ |F_1|^2 ]$.
Let $M$ be a positive integer, 
$\mathbf{c}=(c_1,\ldots,c_M)^T$ be a constant vector, 
$\mathbf{t}=(t_1,\ldots,t_M)^T$ be a non-negative real vector and
$\boldsymbol\Theta(\mathbf{t}) = (\Theta(t_1)-\Theta(0),\ldots,\Theta(t_M)-\Theta(0))^T$ 
where $\Theta(t)$ is defined in (\ref{eq:Thetat}).
We have
\begin{align}
& \mathbb{E}\left[ \frac{1}{\Delta^M} \idotsint_0^{\Delta} \exp(j \mathbf{c}^T \boldsymbol\Theta(\mathbf{t}) ) d\mathbf{t} \right ] \nonumber \\
&\stackrel{(a)}{=} \frac{1}{\Delta^M} \idotsint_0^{\Delta}  \mathbb{E}\left[ \exp(j \mathbf{c}^T \boldsymbol\Theta(\mathbf{t}) ) \right ] d\mathbf{t} \nonumber \\
&\stackrel{(b)}{=} \frac{1}{\Delta^M} \idotsint_0^{\Delta}  \exp\left(- \frac{1}{2} \mathbf{c}^T \Sigma(\mathbf{t}) \mathbf{c} \right)  d\mathbf{t} \nonumber \\
&\stackrel{(c)}{=} \idotsint_0^{1}  \exp\left(- \frac{\Delta}{2} \mathbf{c}^T \Sigma(\mathbf{t}) \mathbf{c} \right)  d\mathbf{u}
\label{eq:E_FM_formula}
\end{align}
where $d\mathbf{t} = dt_M \ldots dt_1$ and
$\Sigma(\mathbf{t})$ is the covariance matrix of $\boldsymbol\Theta(\mathbf{t})$ whose entries are given by
\begin{align}
\Sigma_{ij}(\mathbf{t}) = 2\pi \beta \min\{t_i,t_j\}
,\text{ for }
i,j = 1,\ldots,M.
\label{eq:Sigma_ij}
\end{align}
Step $(a)$ follows from the linearity of expectation,
$(b)$ follows by using the characteristic function of a Gaussian random vector, and
$(c)$ follows from the transformation of variables $\mathbf{t} = \mathbf{u} ~ \Delta$ .
We define 
\begin{align}
a = e^{-\pi \beta \Delta}
\label{eq:a_def}
\end{align}
and use $M=2$ and $\mathbf{c} = (-1,1)^T$ in (\ref{eq:E_FM_formula}) to compute
\begin{align}
\mathbb{E}[ |F_1|^2 ]
&= 2\frac{a-1-\log{a}}{(\log{a})^2}.
\label{eq:EF2}
\end{align}
We also have, using $M=4$ and $\mathbf{c} = (-1,1,-1,1)^T$ in (\ref{eq:E_FM_formula}),
\begin{align}
&\mathbb{E}[ |F_1|^4 ] \label{eq:EF4}\\
&= \frac{783 - 784 a + a^4 + 540 \log{a} + 240 a \log{a} + 144 (\log{a})^2}{18 (\log{a})^4}. \nonumber
\end{align}
Computing the integrals is tedious but straightforward.
Finally, it follows from (\ref{eq:mean_square_G-1}), and (\ref{eq:a_def})~--~(\ref{eq:EF4}) that
\begin{align}
\lim_{\Delta \rightarrow 0} \frac{\mathbb{E}[ \left(G - 1 \right)^2 ]}{\Delta^2}
=  (\pi \beta)^2 ~
\lim_{a \rightarrow 1} \frac{\mathbb{E}[ \left(G - 1 \right)^2 ]}{(\log{a})^2}
=
\frac{(\pi \beta)^2}{9}.
\end{align}

\bibliographystyle{unsrt}
\bibliography{ref8}

\end{document}